\documentclass[preprint,nofootinbib,aps,showpacs,showkeys]{revtex4}
\usepackage{graphpap}
\usepackage{amsmath}
\usepackage{amsfonts,amssymb}
\usepackage{graphicx}

\begin{document}
\title{ \Large Linked-cluster Tamm-Dancoff Field Theory}

\author{Amir H. Rezaeian}
\email{Rezaeian@Theory.phy.umist.ac.uk}
\author{Niels R. Walet}
\email{Niels.Walet@umist.ac.uk} \affiliation{Department of
Physics, UMIST, PO Box 88, Manchester, M60 1QD, UK}

\begin{abstract}
To solve the relativistic bound-state problem one needs to
systematically and simultaneously decouple the high-energy from
the low-energy modes and the many-body from the few-particle
states using a consistent renormalization scheme. In a recent
paper we have shown that one such approach can be a combination of the
coupled cluster method as used in many-body theory and the
Wilsonian exact renormalization group. Even though the method is
intrinsically non-perturbative, one can easily implement a loop
expansion within it. In this letter we provide further support for
this aspect of our formalism by obtaining results for the two-loop
renormalized $\phi^{4}$ theory. We show that the non-unitary
representation inherent in our method leads to an economic
computation and does not produce any non-hermiticity in the relevant terms.
\end{abstract}
\maketitle
\newpage

The Tamm-Dancoff approximation \cite{tamm} was developed in the
1950's to describe a relativistic bound state in terms of a small
number of particles.  It was soon revealed that the Tamm-Dancoff
truncation gives rise to a new class of non-perturbative
divergences, since the truncation does not allow us to take into
account all diagrams at a given order in perturbation theory. On
the other hand, any naive renormalization violates Poincar\'{e}
symmetry and the cluster decomposition property. Two very
different remedies for this issue are the use of light-front
dynamics (see, e.g. \cite{light}) and the application of the
coupled cluster method (CCM) \cite{ccm}.  However, both methods
are too complicated to attack the issues in a self-consistent way.
More recently, Glazek and Wilson \cite{glazek} and independently
Wegner \cite{wegner} introduced an elegant approach for such
problems, the so-called similarity renormalization group (SRG).
There are several problems  with this approach: The Hamiltonian can
not be systematically diagonalized in particle number space, the
computation is extremely complex and there is no efficient
non-perturbative scheme.

In Ref. \cite{me}, we have proposed a new formulation for the
Tamm-Dancoff renormalization in the context of the CCM
\cite{bishop1, bishop} by utilising the Wilsonian exact
renormalization group \cite{ERG}. Our method resembles the SRG
approach, since we employ a similarity transformation to decouple
the high-energy from the low-energy modes, leading to partial
diagonalization of the Hamiltonian. Since we apply a double
similarity transformation using linked-cluster amplitudes, we
produce a biorthogonal representation of the problem, which is not
necessarily unitary. There is a long tradition of such approaches
in nuclear many-body theory \cite{su}. One of our goals in this
letter is to show that the non-hermiticity is not a problem, and
leads to an economic computation. The method is intrinsically
non-perturbative (it can be conceived as a topological
expansion in the number of correlated excitations), but one can
implement a loop expansion within it, and we shall concentrate on
that aspect in this letter. By construction, we design a transformation
which does not produce any fake divergences due to ``small energy
denominators'' which plague old-fashioned perturbation theory in
the Hamiltonian approach \cite{old}. The Poincar\'{e} invariance
and the cluster decomposition property can be maintained in
principle at every level of truncation regardless of
regularization scheme, by requiring a set of decoupling conditions
\cite{me}.

In this letter we supply further evidence for the efficiency of
our method by computing the 2-loop renormalization of $\phi^4$
theory. We only include a succinct discussion of the key elements
of the mathematical framework; details can be found in Ref.
\cite{me}. Notice that our formulation does not depend on the form
of dynamics, i.e. on the choice of quantization (hyper-)plane.

 Consider a system described by a Hamiltonian $H(\Lambda)$
which has, at the outset, a large cut-off $\Lambda$. We assume
that generally the renormalized Hamiltonian
$H^{\text{eff}}(\lambda)$ up to scale $\lambda$ can be expressed
as
\begin{equation}
H^{\text{eff}}(\lambda)=H(\lambda)+H_{C}(\lambda), \label{oh}
\end{equation}
where $H_{C}(\lambda)$ is a {}``counterterm''. Our aim is to
construct the renormalized Hamiltonian by obtaining this
counterterm. We define two subspaces, the model-space
$P:\{|L\rangle \bigotimes|0,b\rangle_{h}, L\leq \mu\}$ and the
complement-space $Q: \{|L\rangle \bigotimes
\left(|H\rangle-|0,b\rangle_{h}\right),\mu<H\leq\Lambda \}$. The
ket $|0,b\rangle_{h}$ is the bare high energy vacuum (the ground
state of the free high-momentum Hamiltonian). The $P$-space
contains interacting low-energy states and $Q$-space contains the
orthogonal complement (the symbols $|L\rangle$ and $|H\rangle$
denote generic low- and high-energy states respectively). Our
renormalization approach is based on decoupling of the complement
space $Q$ from the model space $P$ by using a non-unitary
transformation. The transformation of $H(\Lambda)$ is defined by
\begin{equation}
\overline {H}=e^{\hat{S}'(\mu,\Lambda)}e^{-\hat{S}(\mu,\Lambda)}H(\Lambda)e^{\hat{S}(\mu,\Lambda)}e^{-\hat{S}'(\mu,\Lambda)}\equiv H(\mu)+\delta H(\mu,\Lambda),
\end{equation}
where the operator $\hat{S}(\hat{S}')$ is a functional of certain
mapping operators between $P$- and $Q$-space \cite{me}. By means
of this transformation one may identify the effective interaction
$\delta H(\mu,\Lambda)$ containing the physics between the scale
$\Lambda$ and $\mu$.  According to our prescription we expand
$\hat{S}(\hat{S}')$ in terms of independent coupled cluster
excitation $I$,
\begin{eqnarray}
&&\hat{S}=\sum_{m=0}\hat{S}_{m}\left(\frac{\mu}{\Lambda}\right)^{m}, \hspace{2cm} \hat{S}_{m}=\sideset{}{'}\sum_{I}\hat{s}_{I}C^{\dag}_{I},\nonumber\\
&&\hat{S}'=\sum_{m=0}\hat{S}'_{m}\left(\frac{\mu}{\Lambda}\right)^{m}, \hspace{2cm} \hat{S}'_{m}=\sideset{}{'}\sum_{I}{}\hat{s}'_{I}C_{I}.\label{s}\
\end{eqnarray}
Here the primed sum means that $I\neq 0$, and momentum
conservation is included in $\hat{s}_{I}$ and $\hat{s}'_{I}$. The
$C_{I}$ and $C^{\dag}_{I}$ are annihilation and creation operators
in the high-energy Fock space for a given quantization scheme (e.g., equal time or light-cone).
The indices ${I}$ define a subsystem, or cluster within the full
system of a given configuration. Notice, that the choice of the
operators $\hat{S}(\hat{S}')$ is not generally unique, due to
Haag's theorem \cite{haag}. This ambiguity corresponds to the
possibility of the choice of a different but equivalent
representation of the canonical variables. However the physical
quantities remain invariant under change of operators
$\hat{S}(\hat{S}')$.

It is well-known in many-body applications that the
parametrization Eq.~(\ref{s}) has the following properties: 1) it
satisfies proper size-extensivity and conforms with the Goldstone
linked-cluster theorem at any level of approximation 2) it is compatible
with Hellmann-Feynman theorem, 3) the phase space $
\{\hat{s}_{I},\hat{s}'_{I}\}$ for a given $m$ is a symplectic
differentiable manifold. These features are related to each other and
one can not give up one without losing the others as well
\cite{bishop}.  The price we pay for these desirable features is that
the representation is not longer hermitian.

One may now impose the decoupling conditions, leading to
diagonalization of the transformed Hamiltonian matrix,
\begin{eqnarray}
&&Q\overline{H}P=0 \to \langle 0|C_{I}e^{-\hat{S}}He^{\hat{S}}|0\rangle=0, \label{eq17}\\
&&P\overline{H}Q=0 \to \langle 0|e^{\hat{S}'}e^{-\hat{S}}He^{\hat{S}}e^{-\hat{S}'}C^{\dag}_{I}|0\rangle=0,\label{eq19}\
 \end{eqnarray}
where $I\neq 0$. One can show that the right hand side of Eq.~(\ref{eq17},\ref{eq19}) are derivable from the dynamics of the quantum system \cite{me}. Thus the effective low-energy Hamiltonian is
\begin{equation}
\hat{H}^{\text{eff}}=P\bar{H}P \equiv ~_{h}\langle
b,0|e^{\hat{S}'(\mu,\Lambda)}e^{-\hat{S}(\mu,\Lambda)}H(\Lambda)e^{\hat{S}(\mu,\Lambda)}e^{-\hat{S}'(\mu,\Lambda)}|0,b\rangle_{h}.
\label{eq15}
\end{equation}
 One may prove that the effective low-energy operators
 Eq.~(\ref{eq15}) supplemented by decoupling conditions
 Eq.~(\ref{eq17},\ref{eq19}) indeed have the same low-energy
 eigenvalues as the original Hamiltonian \cite{me}\footnote{It is of interest
 that according to our approach, various effective low-energy
 Hamiltonians can be constructed without invoking perturbation theory
 or hermiticity.}. The states in the full Hilbert space are
 constructed by adding clusters of high-energy correlation to states in
 the $P$-space, or equivalently by integrating out the high-energy modes from
 the Hamiltonian. It is immediately clear that states in interacting
 Hilbert space are normalized, due to the nature of a similarity
 transformation.

The decoupling property makes the $P$ sector of the
 truncated Fock space independent of the rest. This means that the
 contribution of the excluded sector is taken into account by imposing
 the decoupling conditions. One can show that the energy-dependent
 Bloch-Feshbach formalism \cite{b} is thereby made free of the
 small-energy denominators which spoil perturbation theory
\cite{me}. One can then determine the counterterm by requiring
coupling coherence \cite{coh}, namely that the transformed
Hamiltonian Eq.~(\ref{eq15}) has the generic form given in
Eq.~(\ref{oh}), with $\lambda$ replaced by $\mu$. This requirement
must be satisfied on an infinitely long renormalization group
trajectory and thus produces a renormalized Hamiltonian. The
individual amplitudes for a given $m$,
\(\{\hat{s}_{I}^{m},\hat{s}'^{m}_{I}\}
\equiv\{\hat{s}_{I},\hat{s}'_{I}\}_{m}\), have to be fixed by the
dynamics of a quantum system incorporated the decoupling
conditions Eq.~(\ref{eq17},\ref{eq19}). Equations
(\ref{eq17},\ref{eq19}) provide two sets of exact, microscopic,
operatorial coupled non-linear equations for the ket and bra
states which describes the flow of the coefficients in
Eq.~(\ref{s}).  One can solve the coupled equations in
Eq.~(\ref{eq17}) to obtain $\{\hat{s}_{I}\}_{m}$ and then use them
as an input in (\ref{eq19}). The conditions given in
Eq.~(\ref{eq17}) and (\ref{eq19}) imply that all
interactions of high-momentum particles should be removed from the
transformed Hamiltonian $\hat{H}^{\text{eff}}(\mu)$ in
Eq.~(\ref{eq15}). These are sufficient requirements to ensure
partial diagonalization of the Hamiltonian in particle and
momentum space. 

In practice one needs to truncate both sets of coefficients
$\{\hat{s}_{I},\hat{s}'_{I}\}_{m}$ at a given order of $m$ in the
ratio of cutoffs. A consistent truncation scheme is the so-called
SUB($n,m$) scheme, where we include up to $n$-body operators
$\{\hat{S},\hat{S}'\}$ and truncate the expansion in ($\mu/\Lambda$)
at order $m$. The choice of $n$ depends on the Hamiltonian interaction
and needs to be fixed from the outset.

We now apply this formalism to the computation of the effective
Hamiltonian for $\phi^{4}$ theory up to two-loop order in
equal-time quantization. The bare $\phi^{4}$ theory Hamiltonian is
\cite{ph4}
\begin{equation}\label{a1}
H=\int d^{3}x\left(
\frac{1}{2}\pi^{2}(x)+\frac{1}{2}\phi(x)\big(-\nabla^{2}+m^{2}\big)\phi(x)+g\phi^{4}(x)\right).
\end{equation}
According to our logic the ultraviolet-finite Hamiltonian is
obtained by introducing counterterms, which depend on the UV
cutoff $\Lambda$ and some arbitrary renormalization scale. This
redefines the parameters of the theory and defines the effective
low-energy Hamiltonian. The renormalized Hamiltonian has the form
\begin{equation}
H=\int d^{3}x\left(
\frac{Z_{\pi}}{2}\pi^{2}(x)+\frac{1}{2}\sqrt{Z_{\phi}}\phi(x)\big(-\nabla^{2}+Z_{m}m^{2}\big)\sqrt{Z_{\phi}}\phi(x)+Z_{g}Z_{\phi}^{2}g
\phi^{4}(x)+...\right).
\end{equation}
Even though the newly generated interactions are sensitive to the
regularization scheme (as
 is well known \cite{local}, a sharp cutoff may lead to new
non-local interaction terms), nevertheless one can ignore these if
they are irrelevant in the renormalization group sense. We now
split field operators into high- and low-momentum modes;
$\phi(x)=\phi_{L}(x)+\phi_{H}(x)$, where $\phi_{L}(x)$ denotes
modes of low-frequency with momentum $k\leq \mu$ and $\phi_{H}(x)$
denotes modes of high-frequency with momentum constrained to a
shell $\mu<k\leq\Lambda$. The field $\phi_{H}(x)$ can be conceived
as a background to which the $\phi_{L}(x)$-modes are coupled.
Therefore, in the standard diagrammatic language, integrating out
the high-frequency modes $\phi_{H}(x)$ implies that only
high-frequency modes appear in internal lines. The field
$\phi_{H}(x)$ is represented in Fock space as
\begin{equation}
\phi_{H}(x)=\sum_{\mu<k\leq\Lambda}\frac{1}{\sqrt{2\omega_{k}}}(a_{k}e^{ikx}+a^{\dag}_{k}e^{-ikx}),\label{a4}
\end{equation}
where $\omega_{k}=\sqrt{k^{2}+m^{2}}$ and the operators $a_{k}$
and $a_{k}^{\dag}$ satisfy the standard boson commutation rules.
From now on all summations are implicitly over the high-frequency
modes $\mu<k\leq\Lambda$. The Hamiltonian in terms of high- and
low-frequency modes can be written as, after normal ordering with
respect to high-frequency modes,
\begin{equation}
 H=H_{1}+H_{2}+V_{B}+V_{C}+V_{A},
\end{equation}
where we define,
\begin{eqnarray}\label{a5}
 H_{1}&=&\int\left(
\frac{1}{2}\pi_{L}^{2}(x)+\frac{1}{2}\phi_{L}(x)\big(-\nabla^{2}+m^{2}\big)\phi_{L}(x)+g
\phi_{L}^{4}(x)\right),\nonumber\\
H_{2}&=&\sum\omega_{k}a^{\dag}_{k}a_{k},\nonumber\\
V_{B}&=&g\sum\int\frac{e^{i(p+q+r-k)x}}{\sqrt{\omega_{k}\omega_{p}\omega{q}\omega_{r}}}a^{\dag}_{k}a_{p}a_{q}a_{r}
+\frac{3e^{i(p+q-r-k)x}}{4\sqrt{\omega_{k}\omega_{p}\omega{q}\omega_{r}}}a^{\dag}_{k}a^{\dag}_{p}a_{q}a_{r}
\nonumber\\
&&+6\phi_{L}(x)\frac{e^{i(p+q-k)x}}{\sqrt{2\omega_{k}\omega_{p}\omega_{q}}}a^{\dag}_{k}a_{p}a_{q}+
3\big(\phi^{2}_{L}(x)+\frac{1}{2\omega_{r}}\big)\frac{e^{i(k-p)x}}{\sqrt{\omega_{k}\omega_{p}}}a^{\dag}_{p}a_{k}\nonumber\\
&&+\frac{3\phi_{L}^{2}(x)}{2\omega_{r}}+\text{h.c.},\nonumber\\
V_{C}&=&g\sum\int
V^{4}_{C}~a^{\dag}_{k}a^{\dag}_{p}a^{\dag}_{q}a^{\dag}_{r}+V^{3}_{C}
~a^{\dag}_{k}a^{\dag}_{p}a^{\dag}_{q}+V^{2}_{C}~a^{\dag}_{k}a^{\dag}_{p}+V^{1}_{C}~a^{\dag}_{k},\nonumber\\
V_{A}&=&V_{C}^{\dag},\nonumber\\
V^{1}_{C}&=&\Big(\frac{6\phi_{L}(x)}{\omega_{p}}+4\phi_{L}^{3}(x)\Big)\frac{e^{-ikx}}
{\sqrt{2\omega_{k}}},\hspace{2.5cm}
V^{2}_{C}=3\Big(\phi^{2}_{L}(x)+\frac{1}{2\omega_{r}}\Big)\frac{e^{-i(k+p)x}}{\sqrt{\omega_{k}\omega_{p}}},\nonumber\\
V^{3}_{C}&=&2\phi_{L}(x)\frac{e^{-i(k+p+q)x}}{\sqrt{2\omega_{k}\omega_{p}\omega_{k}}},\hspace{3.7cm}
V^{4}_{C}=\frac{e^{-i(k+p+q+r)x}}{4\sqrt{\omega_{k}\omega_{p}\omega_{q}\omega_{r}}}.\
\end{eqnarray}
The high-energy configurations in the Fock space are
specified by
\(\{C_{I}\to {\tiny \prod_{i=1}} a_{k_{i}}\}\) and \(\{C^{\dag}_{I}\to
 \prod_{i=1}a_{k_{i}}^{\dag})\}\). Up to two-loop expansion, our
 renormalization scheme requires to keep $S(S')$ at least to order $n=4$,
 which allows us to eliminate the pure terms $V_{C}$ and $V_{A}$ at a lower level of expansion. The
 $\hat{S}(\hat{S})$ operators consistent with a $SUB(4,m)$ truncation
 scheme are,
\begin{eqnarray}
\hat{S}_{m}&=&\int\sum\left(\hat{S}^{1}_{m}~a^{\dag}_{k}+\hat{S}^{2}_{m}~a^{\dag}_{k}a^{\dag}_{p}+ \hat{S}^{3}_{m}~a^{\dag}_{k}a^{\dag}_{p}a^{\dag}_{q}+\hat{S}^{4}_{m}~a^{\dag}_{k}a^{\dag}_{p}a^{\dag}_{q}a^{\dag}_{r}\right),\nonumber\\
\hat{S}'_{m}&=& \int\sum \left(\hat{S}'^{1}_{m}~a_{k}+\hat{S}'^{2}_{m}~a_{k}a_{p}+\hat{S}'^{3}_{m}~a_{k}a_{p}a^{\dag}_{q}+\hat{S}^{4}_{m}~a_{k}a_{p}a_{q}
a_{r}\right).\label{a6}\
\end{eqnarray}
One can expand Eqs.~(\ref{eq17},\ref{eq19}) in terms of $\mu/\Lambda$,
leading to the introduction of a consistent hierarchy of equations in
powers of $m$ \cite{me}, which can be solved for the coefficients
$\hat{S}_{m}(\hat{S}'_{m})$ in Eq.~(\ref{a6}). We split the diagonalization of the
Hamiltonian matrix in an upper and lower triangle part, using the
double similarity transformation. One may notice that the ``most
non-diagonal'' terms in the Hamiltonian are $V_{C}$ and $V_{A}$ (in
the light-front Hamiltonian such terms do not exist because modes with
longitudinal momentum identically zero are not allowed). The potential
$V_{B}$ is already partially diagonalized and does not change the
vacuum of the high-energy states.  Therefore, here we employ a minimal
scheme, aiming at removal of $V_{A}$ and $V_{C}$ only.

We restrict ourselves to the elimination
of the high-energy degrees of freedom up to the first order in the
coupling constant $g$ and second order in the ratio of cutoffs
\(\mu/\Lambda\). Therefore, our truncation scheme is called
$SUB(4,2)$. For $m=0$ one finds,
\begin{eqnarray}
  S^{1}_{0}&=&-g\frac{V^{1}_{C}}{\omega_{k}}, \hspace{4.8cm}
  S^{2}_{0}=-g\frac{V^{2}_{C}}{\omega_{k}+\omega_{p}},\nonumber\\
  S^{3}_{0}&=&-g\frac{V^{3}_{C}}{\omega_{k}+\omega_{p}+
  \omega_{q}}, \hspace{3cm}
S^{4}_{0}=-g\frac{V^{4}_{C}}{\omega_{k}+\omega_{p}+
  \omega_{q}+\omega_{r}},\label{a7}\
\end{eqnarray}
where the $V^{1-4}_{C}$ are defined in Eq.~(\ref{a5}). Here, one
has $S'_{0}=S_{0}^{\dag}$ \cite{me}. At this stage the results for
the one-loop renormalization can be computed. We evaluate the
effective Hamiltonian by substituting $S(S')$ from Eqs.~(\ref{a6})
and (\ref{a7}) into Eq.~(\ref{eq15}). In order to achieve
renormalization, one should identify the potentially divergent
terms ( when $ \Lambda\to \infty$) in the expansion of
$H^{\text{eff}}(\mu)$. Such a process generally can be done by
inventing a power-counting rule, using the property
$S_{n}\simeq\frac{\mu}{\Lambda }S_{n-1} $. Here we take
$\omega_{k}\simeq|k|$ for $\mu\gg m$ and replace \(\sum_{k}\) by
\(\int \frac{d^{3}k}{(2\pi)^{3}}\). The standard tadpole one-loop
mass renormalization arises from $V_{B}$ due to normal-ordering.
We add this divergent term to $H_{1}$ and renormalize the bare
mass
\begin{eqnarray}
\delta H^{\text{1-loop}}&=& \langle 0|V_{B}|0\rangle=6g\sum\int \frac{\phi^{2}(x)}{2\omega_{k}}=\frac{3g}{4\pi^{2}}(\Lambda^{2}-\mu^{2})\int d^{3}x \phi^{2}(x),\nonumber\\
Z_{m}&=& 1-\frac{3g}{2\pi^{2}}(\Lambda^{2}-\mu^{2}).\
\end{eqnarray}
In this order the contribution of the terms \([V_{C},S],
[V_{A},S']\) and \([H_{1},S(S')]\) are zero, after projection on
to the high-energy vacuum. The only divergent contributions come
from \([V_{A}^{2(3)},S^{2(3)}_{0}]\) due to a double and third
contraction of the high-frequency fields respectively. There are
two other divergent terms, \(([V_{C}^{2(3)},S'^{2(3)}_{0}]\),
however they are harmless and are cancelled out by the divergence
of \([[H_{2},S_{0}],S'^{2(3)}_{0}]\). One thus obtains,
\begin{eqnarray}
\delta H&=& -\frac{18g^{2}}{(2\pi)^{6}}\int\frac{\phi^{2}(x)\phi^{2}(y)}{\omega_{k}\omega_{p}(\omega_{k}+\omega_{p})}e^{i(k+p)(x-y)}\nonumber\\
&-&\frac{12g^{2}}{(2\pi)^{9}}\int\frac{\phi(x)\phi(y)}{\omega_{k}\omega_{p}\omega_{q}(\omega_{k}+\omega_{p}+\omega_{q})}e^{i(k+p+q)(x-y)}.\label{integral}\
\end{eqnarray}
In general evaluation of integrals like Eq.~(\ref{integral}) may
produce non-localities. This is due to the fact that the total
momentum in integrands of Eq.~(\ref{integral}), namely $r_{1}=p+q$
and $r_{2}=k+p+q $ are in the low-momentum space. To evaluate such
integrations, one can firstly reduce the potential divergent
integrals by a change of variable, for example for the first
integrand we use $p,q\to p,r_{1}$, and then expand the integrand
in $r_{1}/p$. Therefore, after expansion and evaluating the
momentum integrals, one may be faced with non-analytic terms in
the low-momentum space. However here these are irrelevant and will
thus be ignored. We find
\begin{eqnarray}
\delta H^{\text{1-loop}}&=&-\frac{9g^{2}}{2\pi^{2}}\ln\left(\frac{\Lambda}{\mu}\right)\int d^{3}x \phi^{4}(x)-\frac{3g^{2}}{2\pi^{4}}(2\ln 2-1)\Lambda^{2}\int d^{3}x \phi^{2}(x)\nonumber\\
&+&\frac{3g^{2}}{16\pi^{4}}\ln\left(\frac{\Lambda}{\mu}\right)\int d^{3}x(\nabla\phi(x))^{2}+\text{finite terms}.
\
\end{eqnarray}
One can immediately deduce the renormalization factors $Z_{g}$ and $Z_{\phi}$ from above expression
\begin{eqnarray}
Z_{g}&=&1+\frac{9g^{2}}{2\pi^{2}}\ln\left(\frac{\Lambda}{\mu}\right),\\
Z_{\phi}&=&1-\frac{3g^{2}}{8\pi^{4}}\ln\left(\frac{\Lambda}{\mu}\right).\label{z}\
\end{eqnarray}
The unknown coefficients in expression $S_{1}$  is computed by
making use of Eq.~(\ref{a7}) and solving coupled equations
(\ref{eq17}), therefore one may yield,
\begin{eqnarray}
S^{1}_{1}&=&\frac{6g
e^{-ikx}}{\omega_{k}^{2}\sqrt{2\omega_{k}}}\Big(
2\phi_{L}(x)-2i\pi_{L}(x)\phi_{L}^{2}(x)-\frac{i\pi_{L}(x)}{\omega_{p}}\Big)-\frac{g}{\omega_{k}}\sum_{\nu=1}^{3}\frac{1}{\nu!}V_{A}^{\nu}S_{1}^{\nu+1},\nonumber\\
S^{2}_{1}&=&\frac{3g
e^{-i(k+p)x}}{(\omega_{k}+\omega_{p})^{2}\sqrt{\omega_{k}\omega_{p}}}\Big(1-i2\pi_{L}(x)
\phi_{L}(x)\Big)-\frac{g}{\omega_{k}+\omega_{p}}\Big([V^{1}_{C},S^{1}_{1}]+ \sum_{\nu=1}^{2}\frac{1}{\nu!}V_{A}^{\nu}S_{1}^{\nu+2}\Big)\nonumber\\
S^{3}_{1}&=&-\frac{2ig
e^{-i(k+p+q)x}}{(\omega_{k}+\omega_{p}+
\omega_{q})^{2}\sqrt{2\omega_{k}\omega_{p}\omega_{q}}}\pi_{L}(x)-\frac{g}
{(\omega_{k}+\omega_{p}+\omega_{q})}\Big(V^{1}_{A}S^{4}_{1}
+\sum_{\nu=1}^{2}[V^{\nu}_{C},S^{3-\nu}_{1}]\Big),\nonumber\\
S^{4}_{1}&=&-\frac{g}{(\omega_{k}+\omega_{p}+\omega_{q}+\omega_{r})}\sum_{\nu=1}^{3}[V^{4-\nu}_{C},S^{\nu}_{1}].\label{a8}\
\end{eqnarray}
In the above expression summation over dummy momentum indices is
assumed. One can find $\hat{S}'_{1}$ in the same manner by
exploiting  Eq.~(\ref{eq19}) and using Eq.~(\ref{a8}) as an input,
which leads to
\begin{equation}
S'^{\nu}_{1}=(S^{\nu}_{1})^{\dag}+S'^{\nu a}_{1} \hspace{2cm} \nu=1,...,4,\label{s'}
\end{equation}
with the notations,
\begin{eqnarray}
S'^{1a}_{1}&=&\frac{g}{\omega_{k}}\Big(\sum_{\nu=1}^{3}\frac{1}{\nu!}S'^{(\nu+1)a}_{1}V^{\nu}_{C}-\sum_{\nu=1}^{3}\frac{1}{\nu!}V^{\nu+1}_{A}S^{\nu}_{1}\Big),\nonumber\\
S'^{2a}_{1}
&=&\frac{g}{\omega_{k}+\omega_{p}}\Big(\sum_{\nu=1}^{2}\frac{1}{\nu!}S'^{(\nu+2)a}_{1}V^{\nu}_{C}(q)-\sum_{\nu=1}^{2}\frac{1}{\nu!}V^{\nu+2}_{A}S^{\nu}_{1}+[V^{1}_{A},S'^{1a}_{1}]\Big),\nonumber\\
S'^{3a}_{1}&=&\frac{g}{\omega_{k}+\omega_{p}+\omega_{q}}\Big(S'^{4a}_{1}V^{1}_{C}-V^{4}_{A}S^{1}_{1}+\sum_{\nu=1}^{2}
[V^{\nu}_{A},S'^{(3-\nu)a}_{1}]\Big),\nonumber\\
S'^{4a}_{1}&=&\frac{g}{(\omega_{k}+\omega_{p}+\omega_{q}+\omega_{r})}\sum_{\nu=1}^{3}[V^{\nu}_{A},S'^{(4-\nu)a}_{1}].\label{a9}\
\end{eqnarray}
The only divergent contribution up to order $g^{2}$ arises from,
\begin{equation}
\delta H=-\langle 0|[H_{1},S_{1}],S'_{0}]|0\rangle,
\end{equation}
After the evaluation of the leading divergent part, we find that
\begin{equation}
\delta H=-\frac{3g^{2}}{16\pi^{4}}\ln\left(\frac{\Lambda}{\mu}\right)\int d^{3}x  \pi^{2}(x),\label{pi}
\end{equation}
which contributes to the two-loop wave-function renormalization
$Z_{\pi}$. By comparing Eqs.~(\ref{z}) and (\ref{pi}), one may
conclude that $Z_{\pi}=Z_{\phi}^{-1}$, as it should be. To finish
the renormalization up to two-loop order, one should also take
into account the contribution at order $g^{3}$. The divergent
terms at this level originate from
\begin{equation}
\delta H=-\langle 0|\Big[[\big(V_{A}+1/2V_{C}+V_{B}\big),S_{0}],S'_{0}\Big]|0\rangle.
\end{equation}
 After a straightforward but lengthy computation one can obtain the leading divergent parts,
\begin{equation}
\delta H=\frac{27g^{3}}{2\pi^{4}}\Big[\big[\ln\left(\frac{\Lambda}{\mu}\right)\big]^{2}+\ln\left(\frac{\Lambda}{\mu}\right)\Big]\int d^{3}x  \phi^{4}(x),
\end{equation}
this term should be added to Eq.~(\ref{integral}), therefore one
can immediately deduce the correct total renormalization factor
$Z_{g}$ up to two-loop order,
\begin{equation}
Z_{g}=1+\frac{9g^{2}}{2\pi^{2}}\ln\left(\frac{\Lambda}{\mu}\right)+\frac{g^{3}}{4\pi^{4}}\left(81\left(\ln\left(\frac{\Lambda}{\mu}\right)\right)^{2}-51\ln\left(\frac{\Lambda}{\mu}\right)\right). \label{f1}
\end{equation}
One can now immediately obtain the well-known \cite{ph4} two-loop
$\beta$-function and anomalous dimension by making use of
Eqs.~(\ref{z}, \ref{f1}).

It is important to point out that the diagonalization at first
order in the coupling constant defines a correct low-energy
effective Hamiltonian which is valid up to order $g^{3}$. Having
said that, from Eq.~(\ref{s'}) one can observe that the
non-hermiticity of the $\hat{S}$ operator appears at order $g^{2}$
and in a lower order of $\mu/\Lambda$. As we have shown,
non-hermiticity is negligible up to two-loop order (asymmetric
terms appear in irrelevant contributions). We conjecture that,
for the present model, non-hermitian terms only appear in irrelevant
contributions, whatever the order of truncation. 

One should note that Eqs.~(\ref{eq17}, \ref{eq19}) are not fully
consistent with the Hellmann-Feynman theorem \cite{me}, although this
choice considerably simplifies the diagonalization of the Hamiltonian
operator. This potentially leads to a different truncation scheme for
the renormalization of other operators. The decoupling requirements in
the exact form Eq.~(\ref{eq17},\ref{eq19}) preserve Poincar\'{e}
symmetry and the cluster decomposition property regardless of
regularization used \cite{me}. Since this is a continuous symmetry,
its preservation leads to an infinite set of constraints on the phase
space, which has been coded in the decoupling equations. Hence, it is
of interest to consider the sensitivity of the Poincar\'{e} symmetry
with respect to a given truncation scheme in the light-front dynamics.

In this letter we have employed a sharp cutoff, however this
idealization should be removed since generally it may lead to
pathologies in renormalization, since it induces non-locality and
moreover potentially violates the gauge symmetry. We have only
investigated a perturbative approach to the problem. Our method is
in principle non-perturbative; this aspect remains to be
exploited.

One of the authors (AHR) acknowledges support from British Government
ORS award and UMIST grant. The work of NRW is supported by the UK
engineering and physical sciences research council under grant GR/N15672.


\begin{thebibliography}{99}
\bibitem{tamm}
I. Tamm, J. Phys. (USSR) {\bf 9}, 499 (1945); S. M. Dancoff, Phys. Rev. {\bf 78}, 382 (1950).
\bibitem{light}
R. J. Perry, A. Harindranath and K. G. Wilson,
Phys. Rev. Lett. {\bf 65}, 2959 (1990); R. J. Perry and
A. Harindranath, Phys. Rev. {\bf D43}, 4051 (1991); K. G. Wilson,
T. S. Walhout, A. Harindranath, W. M. Zhang, R. J. Perry and S. D
Glazek, Phys. Rev. {\bf D49}, 6720 (1994); M. Brisudova and
R. J. Perry, Phys. Rev. {\bf D54}, 1831 (1996).
\bibitem{ccm}
See for example: U. Kaulfuss, Phys. Rev. {\bf D32}, 6, 1421 (1985); G. Hasberg and H. K\"ummel, Phys. Rev. {\bf C33}, 1367 (1986);
H. K\"ummel, Phys. Rev. {\bf D50}, 6556 (1993).
\bibitem{glazek}
S. D. Glazek and K. G. Wilson, Phys. Rev. {\bf D48}, 5863 (1993).
\bibitem{wegner}
F. Wegner, Ann. Physik (Berlin){\bf 3}, 77  (1994).
\bibitem{me}
A. H. Rezaeian and N. R. Walet, Preprint: hep-ph/0212196.
\bibitem{bishop1}
 H. K\"ummel, K. H L\"uhrmann and J. G. Zabolitsky, Phys. Rep. {\bf 36C}, 1 (1978).
\bibitem{bishop}
R. F. Bishop in: {\em Microscopic Quantum many body theories and their applications}, edited by J. Navarro and A. Polls. Lecture notes in Physics {\bf 510}, 1 (1997) and references therein; R. F. Bishop, Theor. Chim. Acta. {\bf 80}, 95 (1991).
\bibitem{ERG}
K. G. Wilson and J. B. Kogut, Phys. Rep. {\bf 12C}, 75 (1974);
C. Bangnuls and C. Bervillier, Phys. Rep. {\bf 348}, 91 (2001).
\bibitem{su} K. Suzuki and S. Y. Lee, Prog. Theor. Phys. {\bf 64}, 209
(1980); H. K\"ummel, Phys. Rev. {\bf C27}, 765 (1983); K. Suzuki and
R. Okamoto, Prog. Theor. Phys. {\bf 76}, 127 (1986); {\bf 75}, 1388
(1988); {\bf 92}, 1045 (1994).  K. Suzuki, Prog. Theor. Phys. {\bf
87}, 937 (1992); K. Suzuki, R. Okamoto and H. Kumaga, Nucl. Phys. {\bf
A580}, 213 (1994);  J. Arponen, Phys. Rev. {\bf
A55}, 2686 (1997).
\bibitem{old}
K. G. Wilson, Phys. Rev. \textbf{140}, B445 (1965); K. G. Wilson,
Phys. Rev. \textbf{D2}, 1438 (1970).
\bibitem{b}
C. Bloch, Nucl. Phys. {\bf 6}, 329 (1958); C. Bloch and J.
Horowitz, Nucl. Phys. {\bf 8}, 91 (1958); H. Feshbach, Ann. Phys.
(N.Y) \textbf{19}, 287 (1962). 
\bibitem{haag}
See for example: N. N. Bogoliubov, A. A. Logunov, A. I. Oksak and I. T. Todorov.
{\it General Principles of Quantum Field Theory} (Moscow: Nauka, 1987).
\bibitem{coh}
R. J. Perry and K. G. Wilson, Nucl. Phys. \textbf{B403}, 587 (1993);
R. Oehme, K. Sibold and W. Zimmermann, Phys. Lett. \textbf{B147},
115 (1984).
\bibitem{ph4}
C. Itzykson and J. Zuber, {\em Quantum Field Theory} (McGraw-Hill, New York, 1980).
\bibitem{local}
T. R. Morris, Int. J. Mod. Phys. \textbf{A9}, 2411 (1994).
\end{thebibliography}
\end{document}